\documentclass[epj]{svjour}
\usepackage{graphics}
\newcommand{\bc}{\begin{center}}
\newcommand{\ec}{\end{center}}
\newcommand{\be}{\begin{equation}}
\newcommand{\ee}{\end{equation}}
\newcommand{\bea}{\begin{eqnarray}}
\newcommand{\eea}{\end{eqnarray}}
\begin{document}
\title{Effective Field Theory and the Nuclear Many Body Problem}
\author{Thomas Sch\"afer
}                     
\institute{Department of Physics, North Carolina State University,
Raleigh, NC 27695 }
\date{Received: date / Revised version: date}
%
\abstract{
We review many body calculations of the equation of state 
of dilute neutron matter in the context of effective field
theories of the nucleon-nucleon interaction. 
} 
\maketitle
%

\section{Introduction}
\label{sec_intro}

 One of the central problems of nuclear physics is to calculate 
the properties of nuclear matter starting from the two-body 
scattering data and the binding energies of few body bound 
states \cite{Bethe:1971xm,Jackson:1984ha}. The nuclear matter 
problem is notoriously difficult. Some of the problems that 
are often mentioned are

\begin{itemize}
\item the large short-range repulsive core in the nucleon-nucleon
interaction

\item the large scattering length in the $^1S_0$ channel, the 
small binding energy of the deuteron, and the small saturation 
density

\item the need to include three (and possibly four) body forces

\item the need to include non-nucleonic degrees of freedoms, 
such as isobars, mesons, quarks, etc.

\end{itemize}

 Ever since the discovery of QCD the classic nuclear matter problem 
has evolved into the broader question of how the properties of 
nuclear matter are related to the parameters of the QCD, the QCD
scale parameter and the masses of the light quarks. 

 Over the last couple of year much progress has been made in 
understanding these kinds of questions in the case of nuclear 
two and three-body bound states \cite{Weinberg:1990rz}. 
Using effective field theory methods it was shown that 
 
\begin{figure}
\resizebox{0.48\textwidth}{!}{%
\includegraphics{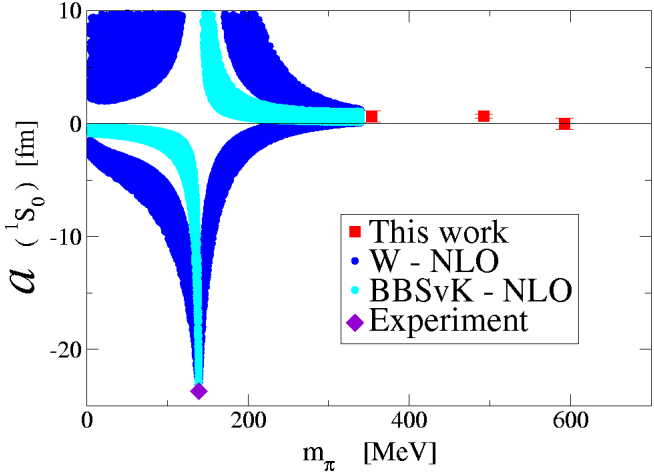}}
\caption{
Quark mass dependence of the scattering length in the $^1S_0$ channel.
The plot shows a combination of unquenched lattice QCD results 
(red points, from Beane et al.~\cite{Beane:2006mx}) and chiral 
extrapolations. The experimental point is shown in purple. Two 
different power counting schemes were employed to constrain
the quark mass dependence, the BBSvK scheme
\protect\cite{Kaplan:1998we,Beane:2001bc} and the W~(Weinberg) 
scheme \protect\cite{Weinberg:1990rz}.}
\label{fig_1s0}
\end{figure}

\begin{itemize}
\item the short range behavior of the nuclear force is not observable.
Using the renormalization group the short distance behavior can be 
modified without changing low energy scattering data and binding energies
\cite{Lepage:1997cs,Bogner:2003wn}.

\item effective field theories can accommodate the large scattering 
lengths in the nucleon-nucleon system \cite{Kaplan:1998we}. The scattering 
lengths depend sensitively on the quark mas\-ses, see Fig.~\ref{fig_1s0}, 
and the large value of $a(^1S_0)$ observed in nature appears to be 
accidental \cite{Beane:2002xf,Epelbaum:2002gb}.

\item  a local three body force is necessary to renormalize the 
two-body force already at leading order. As a consequence, one 
cannot predict three-body binding energies based on two-body 
scattering data alone \cite{Bedaque:1998kg}. 

\item non-nucleonic degrees of freedom, quark effects, relativistic
effects etc.~can be absorbed in local operators. 
\end{itemize}

Effective field theories have also achieved remarkable quantitative
success in describing the available nucleon-nucleon scattering data
below the pion production threshold \cite{Epelbaum:2005pn,Entem:2003ft}.
The long term goal is to achieve a similar qualitative and quantitative
understanding of the nuclear many body problem. 

 In this contribution we shall study a simple limiting case of the 
nuclear matter problem. We shall concentrate on pure neutron matter
at densities significantly below nuclear matter saturation density.
The neutron-neutron scattering length is $a_{nn}=-18$ fm and the 
effective range is $r_{nn}=2.8$ fm. This means that there is a range 
of densities for which the inter-particle spacing is large compared 
to the effective range but small compared to the scattering length.
Neutron matter in this regime exhibits interesting universal properties. 
We are interested in the limit $(k_Fa_{nn})\to\infty$ and $(k_Fr_{nn})
\to 0$, where $k_F$ is the Fermi momentum. From dimensional analysis 
it is clear that the energy per particle at zero temperature has to 
be proportional to energy per particle of a free Fermi gas at the same 
density
\be
\frac{E}{A} = \xi \Big(\frac{E}{A}\Big)_0 = \xi 
\frac{3}{5}\Big(\frac{k_F^2}{2m}\Big).
\ee
The constant $\xi$ is universal, i.~e.~independent of the details of 
the system. Similar universal constants govern the magnitude of the 
gap in units of the Fermi energy and the equation of state at finite
temperature. 

 Universality also implies that the properties of this system can be 
studied using atoms rather than nuclei. The scattering length of 
certain fermionic atoms can be tuned using Feshbach resonances, see 
\cite{Regal:2005} for a review. A small negative scattering length 
corresponds to a weak attractive interaction between the atoms. This 
case is known as the BCS limit. As the strength of the interaction 
increases the scattering length becomes larger. It diverges at the 
point where a bound state is formed. The point $a=\infty$ is called 
the unitarity limit, since the scattering cross section saturates the 
$s$-wave unitarity bound $\sigma=4\pi/k^2$. On the other side of 
the resonance the scattering length is positive. In the BEC limit 
the interaction is strongly attractive and the fermions form deeply 
bound molecules.

\section{Numerical Calculations}
\label{sec_unit_latt}

 The calculation of the dimensionless quantity $\xi$ is a non-perturbative 
problem. In this section we shall describe an approach based on lattice 
field theory methods. The physics of the unitarity limit is captured by 
an effective lagrangian of point-like fermions interacting via a short-range 
interaction. The lagrangian is 
\be 
\label{l_eff}
{\cal L} = \psi^\dagger \left( i\partial_0 
      + \frac{\vec\nabla^2}{2m} \right) \psi 
      - \frac{C_0}{2} \left(\psi^\dagger \psi\right)^2 .
\ee
The standard strategy for dealing with the four-fermion interaction is 
to use a Hubbard-Stratonovich transformation. The partition function can 
be written as \cite{Lee:2004qd}
\be
Z = \int DsDcDc^{\ast} \exp\left[-S\right]  ,
\ee
where $s$ is the Hubbard-Stratonovich field and $c$ is a Grassmann field. 
$S$ is a discretized euclidean action
\bea
S  &=& 
 \sum_{\vec{n},i}\left[  e^{-\hat\mu\alpha_{t}}c_{i}^{\ast}
   (\vec{n})c_{i} (\vec{n}+\hat{0}) \right. \nonumber\\
& & \hspace{0.5cm}\mbox{} \left.
    -e^{\sqrt{-C_0\alpha_{t}}
    s(\vec{n})+\frac{C_0\alpha_{t}}{2}}(1-6h)c_{i}^{\ast}
    (\vec{n})c_{i} (\vec{n})\right] \nonumber\\
& & \mbox{} 
   -h\sum_{\vec{n},l_{s},i}\left[  
    c_{i}^{\ast}(\vec{n})c_{i}(\vec{n}
    +\hat{l}_{s})+c_{i}^{\ast}(\vec{n})c_{i}
  (\vec{n}-\hat{l}_{s})\right]  \nonumber\\
& & \mbox{} 
    +\frac{1}{2}\sum_{\vec{n}}s^{2}(\vec{n}).
\eea
Here $i$ labels spin and $\vec{n}$ labels lattice sites. Spatial and
temporal unit vectors are denoted by $\hat{l}_s$ and $\hat{0}$, 
respectively. The temporal and spatial lattice spacings are $b_\tau$
and $b$. The dimensionless chemical potential is given by $\hat{\mu}
=\mu b_\tau$. We define $\alpha_t$ as the ratio of the temporal and 
spatial lattice spacings and $h=\alpha_t/(2\hat{m})$. Note that for 
$C_0<0$ the action is real and standard Monte Carlo simulations are 
possible. 

\begin{figure}[t]
\resizebox{0.5\textwidth}{!}{%
\includegraphics{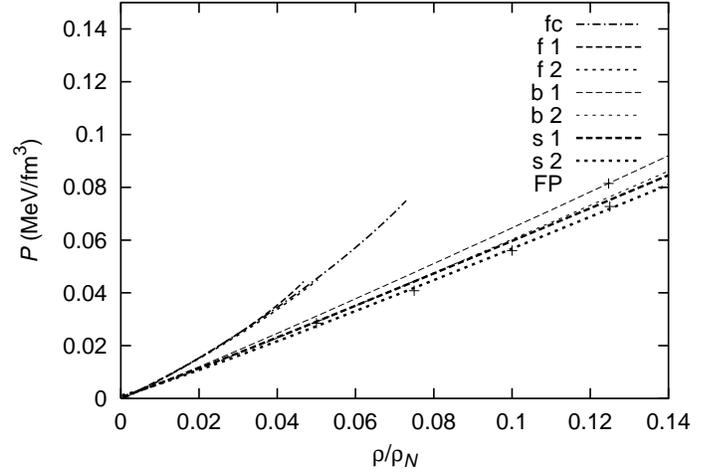}}
\caption{\label{fig_lat_eos}
Equation of state of pure neutron matter at $T=4$ MeV from lattice 
simulations of an effective field theory. Figure taken from 
Lee \& Sch\"afer \cite{Lee:2004qd}. The curves labeled (fc,f1,f2) 
show results for a free gas on the lattice and in the continuum, 
the curves labeled (b1,b2) show ladder sums, and (s1,s2) are 
numerical results on different lattices. We also compare to the 
variational results of Friedman and Pandharipande (FP).}
\end{figure}

 The four-fermion coupling is fixed by computing the sum of all 
two-particle bubbles where on the lattice. Sche\-ma\-ti\-cally, 
\be 
\frac{m}{4\pi a} = \frac{1}{C_0} 
   + \frac{1}{2} \sum_{\vec{p}}\frac{1}{E_{\vec{p}}} ,
\ee
where the sum runs over discrete momenta on the lattice and $E_{\vec{p}}$ 
is the lattice dispersion relation. A detailed discussion of the 
lattice regularized scattering amplitude can be found in 
\cite{Chen:2003vy,Beane:2003da,Lee:2004qd}. For a given scattering 
length $a$ the four-fermion coupling is a function of the lattice 
spacing. The continuum limit correspond to taking the temporal 
and spatial lattice spacings $b_\tau$, $b$ to zero
\be
 b_\tau\mu\to 0, \hspace{1cm} bn^{1/3}\to 0 ,
\ee
where $\mu$ is the chemical potential, $n$ is the density and
$an^{1/3}$ is fixed. We performed numerical simulations at non-zero
temperature and concluded that $\xi=(0.09-0.42)$. Lee studied 
canonical $T=0$ simulations and obtained $\xi=0.25$ \cite{Lee:2005fk}. 
Green Function Monte Carlo calculations give $\xi=0.44$ 
\cite{Carlson:2003wm}, and finite temperature lattice simulations 
have been extrapolated to $T=0$ to yield similar results 
\cite{Bulgac:2005pj,Burovski:2006}.

 Lattice results for the equation of state of dilute neutron matter
at $T=4$ MeV are shown in Fig.~\ref{fig_lat_eos}. For comparison, we 
show variational results obtained by Friedman and Pandharipande
using a phenomenological potential \cite{Friedman:1981qw}. We observe 
that the lattice calculations agree very well with the variational 
result. The pressure is very similar that of non-interacting neutrons 
scaled by a factor $\sim 1/2$. The lattice calculation can be extended 
to higher densities by including explicit pionic degrees of freedom 
in the effective lagrangian \cite{Lee:2004si}. In this case a mild
sign problem returns, but at $T\neq 0$ this sign problem can be handled 
with standard methods.

\section{Analytical Approaches: Large N expansion}
\label{sec_large_n}

 It is clearly desirable to find a systematic analytical approach to 
the dilute Fermi liquid in the unitarity limit. Various possibilities
have been considered, such as an expansion in the number of fermion
species \cite{Furnstahl:2002gt,Nikolic:2006} or the number of spatial 
dimensions \cite{Steele:2000qt,Schafer:2005kg,Nussinov:2004,Nishida:2006br}. 

 We begin with a brief description of the large $N$ approach. The 
physics of the large $N$ limit depends on the symmetries of the 
interaction. One possibility is a $SU(N)$ symmetric interaction 
\cite{Furnstahl:2002gt}
\be 
{\cal L} = \frac{C_0}{2} \left( \psi^\dagger_f\psi_f \right)^2 , 
\ee
where $f=1,\ldots,N$ is a flavor label. A smooth large $N$ limit 
is achieved by keeping $C_0N=c_0$ constant as $N\to\infty$. The 
large $N$ limit is most easily studied by introducing a Hubbard
Stratonovich field $\sigma$ coupled to the density $\psi^\dagger_f
\psi_f$. The leading contribution to the free energy comes from
the free fermion term and the mean field contribution, both of 
which scale as $N$. Subleading $1/N$ corrections arise from 
particle-hole ring diagrams. The problem is that at any fixed
order in the large $N$ expansion the free energy diverges as
the scattering length is taken to infinity. 

 This problem is related to the fact that particle ladders need
to be summed in the unitarity limit. This can be achieved by 
by considering a $Sp(2N)$ symmetric interaction of the form  
\cite{Nikolic:2006}
\be 
{\cal L } = \frac{G_0}{2} (\psi_f {\cal J}^{fg} \psi_g)
 (\psi^\dagger_h {\cal J}^{hi} \psi^\dagger_i),
\ee
where ${\cal J}=(\sigma_2)\otimes \ldots \otimes (\sigma_2)$
and $G_0N=g_0$ constant as $N\to\infty$. This interaction can 
be bosonized using a difermion field $\Phi=(\psi_f{\cal J}^{fg}
\psi_g)/N$. At large $N$ the leading contribution corresponds
to the mean field BCS approximation. The thermodynamic potential 
in the unitarity limit is 
\be 
 \Omega = -N\int\!\frac{d^3p}{(2\pi)^3} \left\{
 \sqrt{\epsilon_p^2+\Phi^2}-\epsilon_p
  -\frac{m\Phi^2}{p^2}\right\}
\ee
with $\epsilon_p=E_p-\mu$. This function can be minimized numerically. 
We find $\Phi_0=1.16\mu$ and $\xi=0.59$. Nikolic and Sachdev studied
$1/N$ corrections near $T_c$ \cite{Nikolic:2006}. These effects are 
not small. They find, for example, $(\mu/T)(T_c)=1.50+2.79/N+O(1/N^2)$. 

\section{Large d expansion}
\label{sec_large_d}

\begin{figure}[t]
\resizebox{0.48\textwidth}{!}{%
\includegraphics{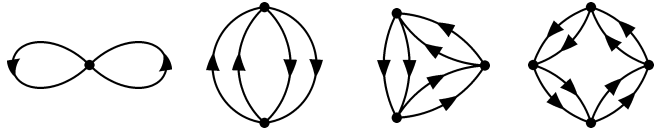}}
\caption{\label{fig_lad}
The particle ladder diagrams shown in this figure give the leading 
order contribution to the ground state energy in the large $d$ 
limit.}
\end{figure}

 Steele suggested that the many body problem of non-relativistic
fermions near the unitarity limit can be studied using an expansion
in $1/d$, where $d$ is the number of spatial dimensions \cite{Steele:2000qt}. 
The main idea is that phase space factors associated with hole lines
are suppressed as $d\to\infty$ so that the leading order contribution
comes from 2-particle ladders, and higher order corrections correspond 
to the hole line expansion of Bethe and Brueckner. 

 Consider the effective lagrangian in equ.~(\ref{l_eff}). We first 
study perturbative corrections to the ground state energy in $d$ 
spatial dimensions. The leading order correction to the energy per
particle is
\be 
\frac{E_1}{A} = \frac{1}{d}  \left[
  \frac{\Omega_{d}C_0k_F^{d-2}M}{(2\pi)^{d}} \right]
 \left(\frac{k_F^2}{2M}\right) .
\ee
This expression indicates that the large $d$ limit should be taken in 
such a way that 
\be
\label{d_scal}
 \lambda \equiv  \left[ 
  \frac{\Omega_{d}C_0k_F^{d-2}M}{d (2\pi)^{d}} \right]
 \stackrel{d\to\infty}{\longrightarrow} {\it const} .
\ee
In the following we wish to study whether this limit is smooth even 
if the theory is non-perturbative. Consider the in medium two-particle 
scattering amplitude in $d$ spatial dimensions. The result is 
\be 
\int\!\! \frac{d^{d}q}{(2\pi)^{d}} \;
  \frac{\theta_q^+}{k^2-q^2+i\epsilon} 
 = f_{vac}(k) + \frac{k_F^{d-2}\Omega_{d}}{2(2\pi)^{d}} 
   f_{PP}^{(d)}(\kappa,s).
\ee
The theta function $\theta_q^+\equiv\theta(k_1-k_F)\theta(k_2-k_F)$
with $\vec{k}_{1,2}=\vec{P}/2\pm \vec{k}$ requires both fermion
momenta to be above the Fermi surface. The first term on the RHS is 
the vacuum contribution. In dimensional regularization the vacuum 
term is purely imaginary and does not contribute to the ground state
energy. The second term is the medium contribution which depends
on the scaled relative momentum $\vec{\kappa}=\vec{k}/k_F$ and 
center-of-mass momentum $\vec{s}=\vec{P}/(2k_F)$. In the large 
$d$ limit we find 
\be 
f_{PP}^{(d)}(s,\kappa) = \frac{1}{d} f_{PP}^{(0)}(s,\kappa)
 \left(1+ O\left(\frac{1}{d}\right)\right),
\ee
which implies that all two-particle ladder diagrams are of the 
same order, see Fig.~\ref{fig_lad}. The sum of all ladder diagrams 
can be calculated by noting that, except for the logarithmic (BCS) 
singularity at $s=0,\kappa=1$, the particle-particle bubble is a 
smooth function of the kinematic variables $s$ and $\kappa$. Hole-hole 
phase space, on the other hand, is strongly peaked at $\bar{s}=\bar{\kappa}=
1/\sqrt{2}$ in the large $d$ limit. We find that $f_{PP}^{(d)}(\bar{s},
\bar{\kappa})=4/d\cdot (1+O(1/d)))$. The ladder sum is a simple geometric 
series and \cite{Schafer:2005kg}
\be
\label{E_pp_d}
 \frac{E}{A} = \left\{ 1 + \frac{\lambda}{1-2\lambda} +
 O\left(\frac{1}{d}\right)\right\}
\left(\frac{k_F^2}{2M}\right),
\ee
where $\lambda$ is the coupling constant defined in equ.~(\ref{d_scal}).
We observe that if the strong coupling limit $\lambda \to \infty$ is taken
after the limit $d\to\infty$ the universal parameter $\xi$ is given 
by 1/2. We have also studied the role of pairing in the large $d$ 
limit. The pairing gap is 
\be 
\Delta = \frac{2e^{-\gamma}E_F}{d} 
  \exp\left(-\frac{1}{d\lambda}\right)
  \left(1+O\left(\frac{1}{d}\right)\right).
\ee
There is no exponential suppression in $d\to\infty$ limit, but 
$\Delta/E_F$ is down by a power of $1/d$. As a consequence the pairing 
energy is sub-leading compared to the result in equ.~(\ref{E_pp_d}). 

\section{Epsilon expansion near four dimensions}
\label{sec_eps}

 Nussinov \& Nussinov observed that the fermion many body system in 
the unitarity limit reduces to a free Fermi gas near $d=2$ spatial 
dimensions, and to a free Bose gas near $d=4$ \cite{Nussinov:2004}. 
Their argument was based on the behavior of the two-body wave function 
as the binding energy goes to zero. For $d=2$ it is well known that 
the limit of zero binding energy corresponds to an arbitrarily weak 
potential. In $d=4$ the two-body wave function at $a=\infty$ has a 
$1/r^2$ behavior and the normalization is concentrated near the origin. 
This suggests the many body system is equivalent to a gas of 
non-interacting bosons.

 A systematic expansion based on the observation of Nussinov \& Nussinov
was studied by Nishida and Son \cite{Nishida:2006br,Nishida:2006eu}. In
this section we shall explain their approach. We begin by restating the 
argument of Nussinov \& Nussinov in the effective field theory language. 
In dimensional regularization $a\to\infty$ corresponds to $C_0\to\infty$. 
The fermion-fermion scattering amplitude is given by
\be 
{\cal A}(p_0,\vec{p})  =
 \frac{ \left(\frac{4\pi}{m}\right)^{\frac{d}{2}}}
      {\Gamma\left(1-\frac{d}{2}\right)}  
   \frac{i}{\left(-p_0+E_p/2-i\delta\right)^{\frac{d}{2}-1}}\; ,
\ee
where $\delta\to 0+$. As a function of $d$ the Gamma function has poles 
at $d=2,4,\ldots$ and the scattering amplitude vanishes at these points. 
Near $d=2$ the scattering amplitude is energy and momentum independent.
For $d=4-\epsilon$ we find
\be
\label{A_4-eps}
{\cal A}(p_0,\vec{p}) =  \frac{8\pi^2\epsilon}{m^2}
 \frac{i}{p_0-E_p/2+i\delta} + O(\epsilon^2) \, .
\ee
We observe that at leading order in $\epsilon$ the scattering amplitude 
looks like the propagator of a boson with mass $2m$. The boson-fermion
coupling is $g^2=(8\pi^2\epsilon)/m^2$ and vanishes as $\epsilon\to 0$. 
This suggests that we can set up a perturbative expansion involving 
fermions of mass $m$ weakly coupled to bosons of mass $2m$. In the 
unitarity limit the Hubbard-Stratonovich transformed lagrangian reads
\be
{\cal L} = \Psi^\dagger\left[
     i\partial_0+\sigma_3\frac{\vec\nabla^2}{2m}\right]\Psi
  + \mu\Psi^\dagger\sigma_3\Psi 
  +\left(\Psi^\dagger\sigma_+\Psi\phi + h.c. \right)\ ,
\ee
where $\Psi=(\psi_\uparrow,\psi_\downarrow^\dagger)^T$ is a two-component 
Nambu-Gorkov field, $\sigma_i$ are Pauli matrices acting in the Nambu-Gorkov 
space and $\sigma_\pm=(\sigma_1\pm i\sigma_2)/2$. In the superfluid phase
$\phi$ acquires an expectation value. We write 
\be
 \phi = \phi_0 + g\varphi, \hspace{1cm}
   g  =\frac{\sqrt{8\pi^2\epsilon}}{m}
       \left(\frac{m\phi_0}{2\pi}\right)^{\epsilon/4} ,
\ee
where $\phi_0=\langle\phi\rangle$. The scale $M^2=m\phi_0/(2\pi)$ was
introduced in order to have a correctly normalized boson field. The scale 
parameter is arbitrary, but this particular choice simplifies some of the 
loop integrals. In order to get a well defined perturbative expansion we 
add and subtract a kinetic term for the boson field to the lagrangian. We
include the kinetic term in the free part of the lagrangian
\bea
{\cal L}_0 &=& \Psi^\dagger\left[i\partial_0+\sigma_3\frac{\vec\nabla^2}{2m}
     + \phi_0(\sigma_{+} +\sigma_{-})\right]\Psi \nonumber  \\
  & & \mbox{}
     + \varphi^\dagger\left(i\partial_0
        + \frac{\vec\nabla^2}{4m}\right)\varphi\, .
\label{l_0}
\eea
The interacting part is 
\bea
{\cal L}_I &=& g\left(\Psi^\dagger\sigma_+\Psi\varphi + h.c\right)
     + \mu\Psi^\dagger\sigma_3\Psi  \nonumber \\
 & & \mbox{}
     - \varphi^\dagger\left(i\partial_0
        + \frac{\vec\nabla^2}{4m}\right)\varphi\, .
\eea
Note that the interacting part generates self energy corrections to 
the boson propagator which, by virtue of equ.~(\ref{A_4-eps}), cancel 
against the kinetic term of boson field. We have also included the 
chemical potential term in ${\cal L}_I$. This is motivated by the fact 
that near $d=4$ the system reduces to a non-interacting Bose gas and
$\mu\to 0$. We will count $\mu$ as a quantity of $O(\epsilon)$. 

\begin{figure}[t]
\resizebox{0.48\textwidth}{!}{%
\includegraphics{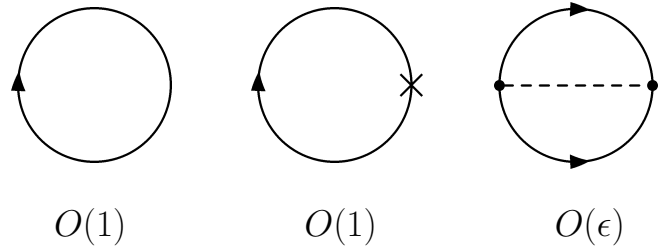}}
\caption{\label{fig_veff_eps}
Leading order contributions to the ground state energy in 
the $\epsilon=4-d$ expansion. Solid lines are fermion propagators, 
dashed lines are boson propagators, and the cross is an insertion
of the chemical potential.}
\end{figure}

 The Feynman rules are quite simple. The fermion and boson propagators 
follow from equ.~(\ref{l_0}) and the fermion-boson vertices are 
$ig\sigma^\pm$. Insertions of the chemical potential are $i\mu\sigma_3$. 
Both $g^2$ and $\mu$ are corrections of order $\epsilon$. In order to 
verify that the $\epsilon$ expansion is well defined we have to check 
that higher order diagrams do not generate powers of $1/\epsilon$. 
Studying the superficial degree of divergence of diagrams one can 
show that there are only a finite number of one-loop diagrams that 
generate $1/\epsilon$ terms. 

 The leading order diagrams that contribute to the effective potential 
are shown in Fig.~\ref{fig_veff_eps}. The first diagram is the free
fermion loop which is $O(1)$. The second diagram is the $\mu$ insertion
which is $O(1)$ because the loop diagram is divergent in $d=4$. The
sum of these two diagrams is
\be
 V_1 =-\int\frac{d^dp}{(2\pi)^d}\, \left\{
   \sqrt{E^2_{\vec p}+\phi_0^2}
    -\frac{\mu E_{\vec p}}{\sqrt{E^2_{\vec p}+\phi_0^2}}
   \right\}\ .
\ee
The integral can be computed analytically. Expanding to first order 
in $\epsilon=4-d$ we get 
\bea
\label{VLOa}
 V_0 &=& \left\{  \frac{\phi_0}{3}\left[ 
   1 + \frac{7-3(\gamma+\log(2))}{6}\, \epsilon \right] \right.
  \nonumber \\ 
     & & \mbox{}\left.
   -\frac{\mu}{\epsilon}\left[ 
   1 + \frac{1-2(\gamma-\log(2))}{4}\, \epsilon \right]\right\}
  \left(\frac{m\phi_0}{2\pi}\right)^{d/2} 
\eea
Nishida and Son also computed the two-loop contribution shown 
in Fig.~\ref{fig_veff_eps}. The result is 
\be 
V_2 = -C\epsilon  \left(\frac{m\phi_0}{2\pi}\right)^{d/2} ,
\ee
where $C\simeq 0.14424$. We can now determine the minimum of the 
effective potential. We find
\be 
\phi_0 = \frac{2\mu}{\epsilon}\,\left[ 1 + 
    (3C-1+\log(2))\,\epsilon + O(\epsilon^2) \right] .
\ee
The value of $V=V_1+V_2$ at $\phi_0$ determines the pressure
and $n=\partial P/\partial \mu$ gives the density. We find
\be 
\label{n_int}
 n = \frac{1}{\epsilon}\,\left[ 1 - \frac{1}{4}
  \left( 2\gamma-1-\log(2) \right)\epsilon \right]  
 \left(\frac{m\phi_0}{2\pi}\right)^{d/2}.
\ee
We can compare this result with the density of a free Fermi gas 
in $d$ dimensions. This equation determines the relation between 
$\epsilon_F\equiv k_F^2/(2m)$ and the density. We get 
\be 
\label{e_f}
\epsilon_F = \frac{2\pi}{m} \left[\frac{n}{2}
  \Gamma\left(\frac{d}{2}+1\right)\right]^{2/d}.
\ee
We determine $\epsilon_F$ for the interacting gas by inserting 
$n$ from equ.~(\ref{n_int}) into equ.~(\ref{e_f}). The universal   
parameter is $\xi=\mu/\epsilon_F$ . We find 
\be 
\label{xi_4-}
\xi = \frac{1}{2}\epsilon^{3/2} 
      + \frac{1}{16}\epsilon^{5/2}\log(\epsilon)
      - 0.025\epsilon^{5/2} + \ldots 
    = 0.475 ,
\ee
where we have set $\epsilon=1$. The calculation has been extended to 
$O(\epsilon^{7/2})$ by Arnold et al.~\cite{Arnold:2006fr}. Unfortunately,
the next term is very large and it appears necessary to combine the 
expansion in $4-\epsilon$ dimensions with a $2+\epsilon$ expansion 
in order to extract useful results. 

\section{Epsilon expansion near two dimensions}
\label{sec_eps_2}

\begin{figure}[t]
\resizebox{0.44\textwidth}{!}{%
\includegraphics{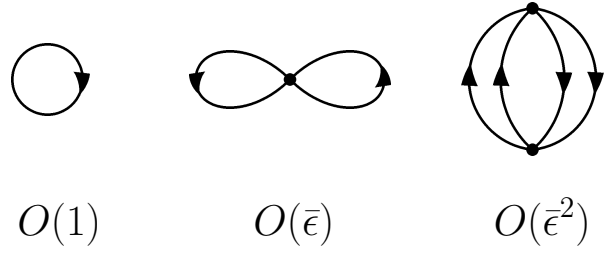}}
\caption{\label{fig_2+}
Leading order contributions to the ground state energy in the 
$\bar\epsilon=d-2$ expansion.}
\end{figure}

 Near two spatial dimensions the scattering amplitude in the 
unitarity limit vanishes linearly in $\bar\epsilon=d-2$
\be 
{\cal A}(p_0,p) = i\frac{2\pi}{m}\, \bar\epsilon + O(\bar\epsilon^2) .
\ee
The coefficient of $\bar\epsilon$ is momentum end energy independent. 
This means that we can set up a perturbative expansion with an 
effective four-fermion coupling $g^2 = 2\pi\epsilon/m$. This expansion
is very similar to the perturbative $(k_Fa)$ expansion studied by
Huang, Lee and Yang \cite{Lee:1957,Huang:1957}, see Fig.~\ref{fig_2+},
but it is not restricted to the weak coupling limit. To $O(\bar
\epsilon)$ the effective potential is given by
\be 
 V_0+V_1 = -P_{free} - \frac{g^2}{4} \rho^2 + O(\bar\epsilon^2), 
\ee
where $P_{free}$ is the pressure of free fermions expanded to $O(\bar
\epsilon)$ and the density is given by 
\be 
\rho = 2\int\!\frac{d^2p}{(2\pi)^2} \Theta (\mu-E_p) .
\ee
From the total pressure we can compute the density and Fermi energy 
as in the previous section. The universal parameter $\xi$ is given by 
\be 
\label{xi_2+}
\xi = 1 - \bar\epsilon + O(\bar\epsilon^2) = 0 \hspace{0.5cm}
  (\bar\epsilon=1).
\ee
Similar to the perturbative expansion pairing is exponentially 
suppressed in the $\bar\epsilon$ expansion. The pairing gap is 
\cite{Nishida:2006eu}
\be 
\phi_0 = \frac{2\mu}{e} 
     \exp\left(-\frac{1}{\bar\epsilon}\right) 
     \left( 1+O(\bar\epsilon) \right),
\ee
which corresponds to the perturbative result of Gorkov and 
Melik-Barkhudarov \cite{Gorkov:1961}. Equation (\ref{xi_2+}) 
shows that the $\bar\epsilon$ expansion is poorly convergent. 
However, the $\bar\epsilon$ expansion is useful in improving 
the convergence of the $\epsilon=d-4$ expansion, and in connecting 
the perturbative $(k_Fa)$ expansion with the physics of the 
unitarity limit.

\section{Outlook}
\label{sec_sum}
 
 In this contribution we focused on an idealized systems 
of neutrons at very low density. The obvious question is 
to what extent these methods can be extended to nuclear 
systems near saturation density. 

 The lattice simulations can easily be extended to include 
finite range effects, explicit pions, and isospin. Some of these
refinements will cause a sign problem in the simulation, but in 
most cases the sign problem can be handled with standard 
methods. A significant amount of work will be required in 
order to reduce discretization errors to the point where the 
interactions are quantitatively reliable all the way up 
to momenta on the order of the Fermi momentum in nuclear 
matter. 
 
 The large $N$, large $d$, or epsilon expansions are easily 
extended to interactions with a finite scattering length and a 
finite effective range \cite{Nikolic:2006,Rupak:2006jj,Chen:2006wx}.
There is also no obvious obstacle to including explicit pion 
degrees of freedom. It will be interesting to extend these 
methods to systems of protons and neutrons. In this case 
three-body forces have to be included. The central question
is whether saturation can be achieved, and whether effective 
field theories provide a qualitative understanding of the 
Coester line \cite{Coester:1970}. 

 We should also note that the many body physics that governs 
the equation of state of nuclear matter near saturation 
density may well be simpler than the physics of the unitarity
limit. Effective range corrections suppress the two-body 
scattering amplitude and nuclear matter is more perturbative 
than dilute neutron matter. As a consequence, perturbative 
calculations using soft potentials or effective interactions
adjusted to nuclear matter properties may well be reliable
 \cite{Bogner:2005sn,Kaiser:2001jx}.

Acknowledgments: The work described in this contribution was 
done in collaboration with S.~Cotanch, C.-W.~Kao, A.~Kryjevski,
D.~Lee and G.~Rupak. This work is supported in part by the 
US Department of Energy grant DE-FG-88ER40388.

\end{document}